\documentclass[11pt,twoside]{article}
\usepackage{newpasp,epsf}
\reversemarginpar

\begin{document}
\title{Acceleration of CR at Large Scale Shocks and Their 
Cosmological Role for Structure Formation in the Universe}
 \author{Francesco Miniati}
\affil{School of Physics and Astronomy, University of Minnesota,
    Minneapolis, MN 55455 USA}
\author{Dongsu Ryu}
\affil{Department of Astronomy \& Space Science, Chungnam National
    University, Daejeon, 305-764 Korea}
 \author{T. W. Jones}
\affil{School of Physics and Astronomy, University of Minnesota,
    Minneapolis, MN 55455 USA}
\author{Hyesung Kang}
\affil{Department of Earth Science, Pusan National University,
    Pusan, 609-735 Korea}

\begin{abstract}
We investigate the dynamical
importance of a newly recognized 
possible source of significant feedback generated during
structure formation; namely cosmic ray (CR) pressure.
We present evidence for the existence of numerous shocks in the 
hot gas of galaxy clusters (GCs).
We employ for the first time an explicit numerical treatment 
of CR acceleration and transport in hydro simulations of structure
formation. According to our results, CRs provide an important fraction 
of the total pressure inside GCs, up to several tenths.
This was true even at high redshift (z=2), meaning that such non-thermal
component could affect the evolution of structure formation.
\end{abstract}

%
%
\section{Introduction}
During the hierarchical process of structure formation,
supersonic gas
infall and merging events invariably generate powerful, 
large and long-lived shock waves (Miniati et al. 1999).
These should produce copious amounts of CRs, by way of diffusive shock
acceleration (e.g. Blandford \& Ostriker 1978),
including both electrons and ions.
In addition, the post-shock gas and diffusively trapped CRs are
mostly advected into
non-expanding regions, such as filaments and clusters. 
It turns out that the energy of most of the CR-protons is only marginally
affected by radiative losses during a Hubble time. 
The important possibility, then, is that
the latter might accumulate inside 
forming structures, storing up a substantial fraction 
of the total pressure there. In addition to cosmic shocks
other sources of CRs are also possible. 
These include AGNs, SNe and stellar winds all of which are
candidates for important contributions to the 
total population of CRs in cosmic structures, although 
they are not discussed here.

There is growing observational evidence that significant 
non-thermal activity takes place in GCs. This evidence 
is provided by extended sources of polarized radio
emission, interpreted as synchrotron radiation from relativistic
electrons (e.g. Hanisch 1984;
Deiss et al. 1997); and by the
detection of radiation in excess to what is
expected from the hot, thermal X-ray emitting Intra 
Cluster Medium (ICM), both in the
extreme ultra-violet (e.g. Lieu et al. 1996;
Kaastra 1998) and in the hard X-ray
band above $\sim 10$ KeV (e.g. Fusco-Femiano et al.
1999; Valinia et al. 1999). Although a coherent picture of the non-thermal
status of the
ICM is still lacking, a very
plausible origin for these radiation excesses is inverse-Compton (IC) 
due to relativistic electrons (e.g. Sarazin \& Lieu 1998).
Based on this assumption and on the measured EUV excess in Coma
cluster, Lieu et al. (1999) have estimated the existence of a CR proton
component
in approximate {\it equipartition} of energy with the thermal gas.

\section{Lessons from Hydro Simulations of Structure Formation}

Fig. 1a illustrates a slice of a typical cosmic structure formed 
in a hydro simulation of structure formation with 
$\Omega_m\equiv \rho_m/\rho_c = 1$, $\sigma_8=1.05$, computational box size
32 $h^{-1}$ Mpc and 256$^3$ cells.
It shows contours of compression ($\nabla \cdot v$)
corresponding to shock waves, superposed on a grayscale image of X-ray
bremsstrahlung emission from the hot ICM (brighter regions correspond to higher
emission). We can easily recognize 
the external, {\it accretion shock waves} enveloping clusters and
filaments and processing for the first time the supersonic (accretion) flows. 
In addition, however, 
it is also shown that the ICM of GCs is
commonly populated by a complex structure of what we call
{\it internal shock waves}. Unlike the external accretion shocks, 
internal shocks propagate
through gas inside formed (or forming)
structures that have already being shock heated.
Such shocks include not only {\it merger shocks}
associated with merger events, 
but also, and more commonly, {\it flow shocks} that are generated 
because of the complexity of the supersonic accretion flows.
They
have similar properties (e.g. size, Mach number) to, but are
more common than the largest merger shocks.
For this reason
they are of primary importance for production of relativistic 
CRs and must be considered when addressing the 
issues on the non-thermal activity inside GCs.

Hydro simulations allow us to estimate roughly the 
expected contribution of cosmic shocks in terms of
CR production over cosmological time-scales.
The relative importance of the CR dynamical role is usually expressed 
as the ratio of the CR pressure to the thermal pressure: $P_{CR}/P_{th}$.
The energy stored in CR can be estimated as a fraction 
$\epsilon_{E_k\rightarrow CR}$
of the total kinetic energy that has 
been processed thorough shocks since a certain epoch, say $z=1.5$, up to
now ($z=0$). Here, $\epsilon_{E_k\rightarrow CR}$ is 
the conversion efficiency of 
kinetic energy into CR energy and can conservatively be 
assumed to be circa 0.1.
Then $P_{CR}/P_{th}$ is given by:
\begin{equation}
\frac{P_{CR}}{P_{th}} = \frac{\epsilon_{E_k\rightarrow CR}}{2~E_{th}(z=0)}\int_{t(z=1.5)}^{t(z=0)} (\Phi_{E_k})_{shock} dt
\end{equation}
where $P_{th}$ is the average pressure in the computational box.
Here, $(\Phi_{E_k})_{shock}$ is the flux of kinetic energy across shocks.
Our conclusions indicate that roughly
$P_{CR}/P_{th} \simeq 6-8 \epsilon_{E_k\rightarrow CR} \sim 0.6-0.8$ 
(see Miniati et al. 1999 for more detail). Thus a large fraction of the
total pressure inside GCs today could be provided by CRs, in rough 
agreement with Lieu et al. (1999).

\section{Preliminary Results from Explicit CR Numerical Treatment}

\begin{figure}
\plotfiddle{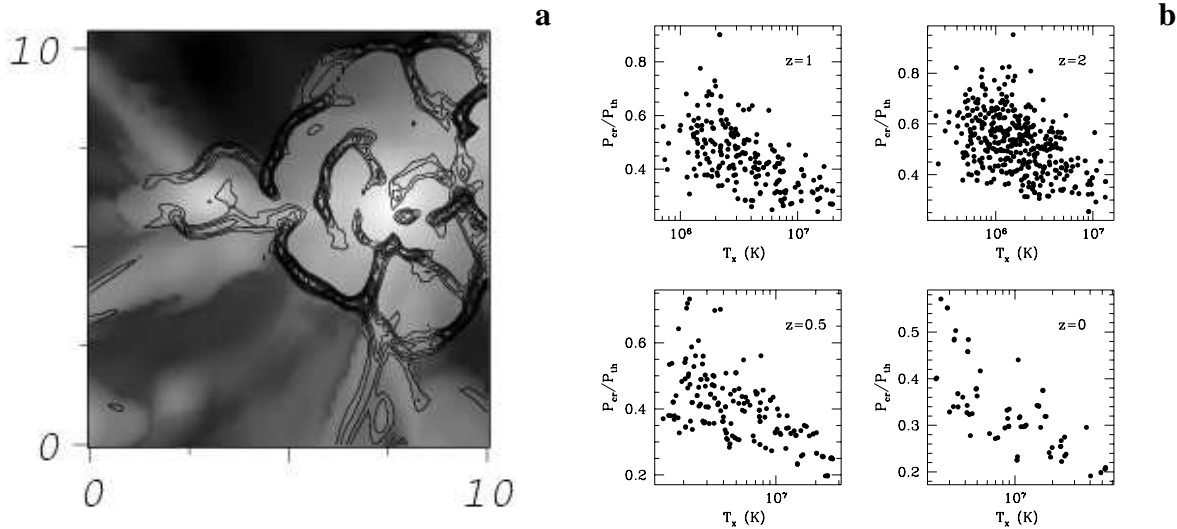}{6.5truecm}{0}{90}{90}{-280}{-440}
\caption{(a) Two dimensional slice of a typical cosmic structure.
Scales are in $h^{-1}$Mpc.
(b) Ratio of CR to thermal pressure as a function of GC temperature at
four different redshifts.}
\end{figure}
We have developed unique numerical tools to treat explicitly 
CR ions acceleration and transport inside
simulations of cosmological models (Jones et al. 1999; Ryu et al. 1999).
With the CR spatial and spectral information provided by our new scheme,
employed in such simulations for the first time, we can assess the 
question of the non-thermal dynamical contribution in GCs more accurately. 
In particular, we can begin to explore
the ratio $P_{CR}/P_{th}$ as a function of the
cluster temperature and for different redshifts. 
In the simulation described here
we have adopted for the fraction of postshock thermal particles to be 
injected at the shock, the value $f_{inj}\simeq 10^{-4}$.
Our results are shown in Fig. 1b. 
First of all, they indicate that for any $T_x$ and $z$
$P_{CR}$ is {\it not} a negligible fraction of $P_{th}$, in accord with 
our previous findings. Also, the four panels show that for any $z$
the ratio $P_{CR}/P_{th}$ tends to be larger for smaller clusters.
Finally, such a ratio not only is still not negligible at high $z$,
but it is actually larger at higher $z$. This indicates
that the evolution
of the large scale structure could be significantly affected by 
this dynamical component. 
\section{Discussion \& Conclusions}

We have shown that the ICM of GCs is commonly populated by 
numerous internal flow shocks with similar characteristic to, but
not necessarily associated with major merger events.
These along with accretion shocks and merger shocks
are likely to play an important role for the non-thermal 
activity of the ICM. We have also shown that CR pressure could
provide a substantial fraction of the total pressure in GCs today,
thus affecting GC mass estimates based on the hydrostatic 
equilibrium assumption and in turn, the baryonic fraction estimates
(which end up being biased high). We have also shown that CR  pressure
was significant already at high $z$, therefore possibly affecting
the evolution of structure formation.
Since this is often used as a tools for 
discriminating among different cosmological models 
(e.g. Carlberg et al. 1997; Bahcall \& Fan 1998), 
the role of CR pressure should be well understood
in order to apply evolutionary arguments with confidence.

\acknowledgments
Support at the University of Minnesota  was provided by NSF and
the U of MN Supercomputing Institute.
FM was supported in part by a Doctoral Dissertation Fellowship at the 
University of Minnesota. 
DR and HK were supported in part by the KOSEF grant 1999-2-113-001-5.

\end{document}